\documentclass[11pt,twoside]{article}
\usepackage{asp2010}
\usepackage{natbib}
\pdfoutput=1
\resetcounters

\aspvolume{**Volume**} % Insert volume number from ASPCS
\aspcpryear{2010} % Insert year received from ASPCS
\aspvoltitle{Asymmetric Planetary Nebula V: The Shaping of Stellar Ejecta} % Insert Full Title in Initial Caps
\aspvolauthor{**Editors, eds.} % Insert Editors -- See next lines

\begin{document}

\title{Proto-Planetary Nebulae with the Spitzer Space Telescope}
\author{Alexa Hart$^{1,2}$, Joe Hora$^2$, Luciano Cerrigone$^3$, Grazia Umana$^4$, Corrado Trigilio$^4$, Martin Cohen$^5$ and Massimo Marengo$^6$}
\affil{$^1$Department of Physics \& Astronomy, University of Denver,\\ 2112 E. Wesley Avenue, Denver, CO 80208, USA}
\affil{$^2$Harvard-Smithsonian Center for Astrophysics, \\160 Garden Street, Cambridge, MA 02138, USA}
\affil{$^3$Max-Planck-Institut fuer Radioastronomie, \\ Auf dem Huegel 69, 53121 Bonn, Germany}
\affil{$^4$INAF-Osservatorio Astrofisico di Catania, \\ Via S. Sofia 75, 95123 Catania, Italy}
\affil{$^5$Astronomy Department, University of California at Berkely, \\ 601 Campbell Hall, Berkeley, CA 94720 USA}
\affil{$^6$Department of Physics \& Astronomy, Iowa State University, \\ A313E Zaffarano, Ames, IA 50011 USA}

\begin{abstract}
The transition from Asymptotic Giant Branch star to Planetary Nebula is short-lived and mysterious. Though it lasts only a few thousand years, it is thought to be the time when the asymmetries observed in subsequent phases arise.  During this epoch, the star is shrouded in thick clouds of dust and molecular gas;  infrared observations are needed to reveal these objects at their most pivotal moment.  I present preliminary results of a Spitzer study of targets spanning the range from post-AGB stars to Planetary Nebulae with the goal of determining the genesis of asymmetry in these objects.

\end{abstract}

\section{The Evolution of Intermediate-Mass Stars}
The life cycle of a star and the manner of its ultimate demise depends on its mass.  Intermediate-mass stars gradually fling their outer layers into space, forming planetary nebulae.  

At the end of the Giant phase, the envelope of intermediate mass stars (defined as  $0.8 \mbox{--}8 M_{\odot}$) expands and the core contracts, resulting in a degenerate C/O core.  This is the beginning of the Asymptotic Giant Branch (AGB) phase; the star will start to burn nuclear fuel in shells surrounding the spent core. During this stage, there is a slow, dense and (presumably) spherically symmetric stream of material leaving the star called the AGB wind.  This wind is interspersed with shells of enhanced density due to pulsation.   These shells have been observed in many Planetary Nebulae such as the Cat's Eye Nebula  (Balick et al., 2001).     
  
As the mass loss process continues,  the thick cloud of dust and gas it creates begins to obscure the star itself in visible wavelengths (Garc{\'{\i}}a-Lario, 2006).  Eventually, the central star will have lost so much mass that the AGB wind will cease; this marks the beginning of the post-AGB phase (Kwok et al. 1993).
Though we can conjecture about what happens next, we have precious little observational evidence for it due to the obscuration of the central star.  However, post-AGB stars observed in the mid-infrared with IRAS apparently fit into an evolutionary sequence with  effective temperatures steadily increasing until the star is hot enough to ionize its envelope (spectral types marching from K, G, F, A to B).  Near the earliest onset of ionization, the star becomes a Transition Object, sometimes called a proto-Planetary Nebula. The transition phase lasts mere thousands of years.

The planetary nebula phase commences when the star ionizes its envelope.  The ionized material is optically thin, so the central star re-emerges in visible wavelengths once the ionization front has traveled sufficiently far out into the envelope.  Morphologies of these young PNe are diverse: bipolar, elliptical, point-symmetric and spherical.  

Though few AGB stars have been spatially resolved, they are generally assumed to have spherically symmetric outflows (as long as they are single).  Yet when the star emerges just a short time later, it is suddenly the central star of a spectacular, and often highly asymmetric, Planetary Nebulae.  Somehow, during this phase of mere thousands of years, the structure of the winds, and perhaps the star itself, change fundamentally.   What happens behind that dark cloud of dust and gas is revealed in infrared observations with the Spitzer Space Telescope.

\section{The Sample}
A sample of targets was chosen to span the entire temporal sequence from star to nebula.  This sequence was divided into three constituent epochs: post-AGB stars, true Transition Objects, and Planetary Nebulae.  Each of these epochs were the subject of an observational program on {\it Spitzer}, divided as follows: 26 post-AGB stars, 36 Transition Objects, and 18 Planetary Nebulae.  

The post-AGB sample was designed to catch stars that are nearing the end of the post-AGB phase, that is, stars that are on the brink of ionizing their envelopes.  To find such stars, we started from targets that were characterized as post-AGB candidates based on their IRAS colors (Suarez et al 2006) and then further constrained them to have a central star of spectral type A and a far-IR excess typical of PNe (Parthasarathy 2000).  The spectral type condition should limit the sample to objects with very hot central stars, which should be those that are at the very end of the post-AGB phase.  Type A stars should not generally be hot enough to ionize surrounding material.

To isolate true Transition Objects, i.e. stars that have just begun to ionize their circumstellar envelopes, a sample of post-AGB candidate (again selected based on IRAS colors and infrared excess) with slightly hotter spectral type B central stars were chosen.  These stars should be just hot enough to emit ionizing radiation.  As an independent measure of ionization, these objects were observed with the VLA; 16 of the 36 were detected, confirming the presence of ionized shells (Umana et al. 2004; Cerrigone et al. 2008). 

The chosen Planetary Nebulae all have H-deficient central stars of late, carbon-rich Wolf-Rayet spectral type ([WC 8--12]).  These PNe may also have  in common a peculiar spectroscopic property: simultaneous evidence for both oxygen-based crystalline silicates and carbon-based PAHs in their dust.  

\section{Dust Chemistry}
An important observational aspect of these stars is their dust chemistry, that is, if the dust around them is Oxygen-rich or Carbon-rich.  This distinction is a consequence of the high binding energy of CO; since the CO molecule will form first, only the more abundant of C or O will be left to form other dust grains.  All AGB stars start out O-rich;  stars with roughly $1.5 \mbox{--}4 M_{\odot}$ can become C-rich via the Third Dredge Up.

 In Carbon-rich sources, we see Polycyclic Aromatic Hydrocarbon (PAH) features at 3.3, 6.2, 7.7, 8.6 and 11.3 $\mu$m$\,$ (Hrivnak et al. 2007), as well as an SiC feature at 11.3 $\mu$m$\,$ (Speck et al. 2009).  Oxygen-rich sources have silicate features at $\sim$10-18 $\mu$m$\,$ and show evidence for water ice.  A small minority of stars show evidence of both PAHs and silicates in their dust (called dual-dust objects), a phenomena which has yet to be understood.  These dual dust spectra could be indicative of either an O-rich environment preserved in a disk while the rest of the envelope transformed to C-rich or simply the formation of PAHs in a C-rich environment as a result of the photo-dissociation of CO molecules. 

Interestingly, Cerrigone et al. 2009 found that approximately $40\%$ of the Transition Object sample were dual-dust objects, a striking contrast to the global occurence of $\sim10\%$ among PNe.  The authors noted that this might be due to a selection effect, in particular the choice of objects with a far-IR excess which may be indicative of the presence of a disk.  Additionally, $70\%$ of those dual-dust objects were radio-detected in free-free emission, establishing an apparent correlation between ionization and the dual-dust phenomenon.  Modeling of the PAH features resulted in the conclusion that the PAH molecules are located in the outflows, far from the central star.  For the detailed analysis, see Cerrigone et al. 2009.

The PNe sample was chosen with this peculiarity in mind:  Cohen et al. 2002 found that four out of six PNe with central stars of type [WCL] in their study were dual-dust objects; the PNe sample was designed to further explore this apparent connection. 

\articlefiguretwo{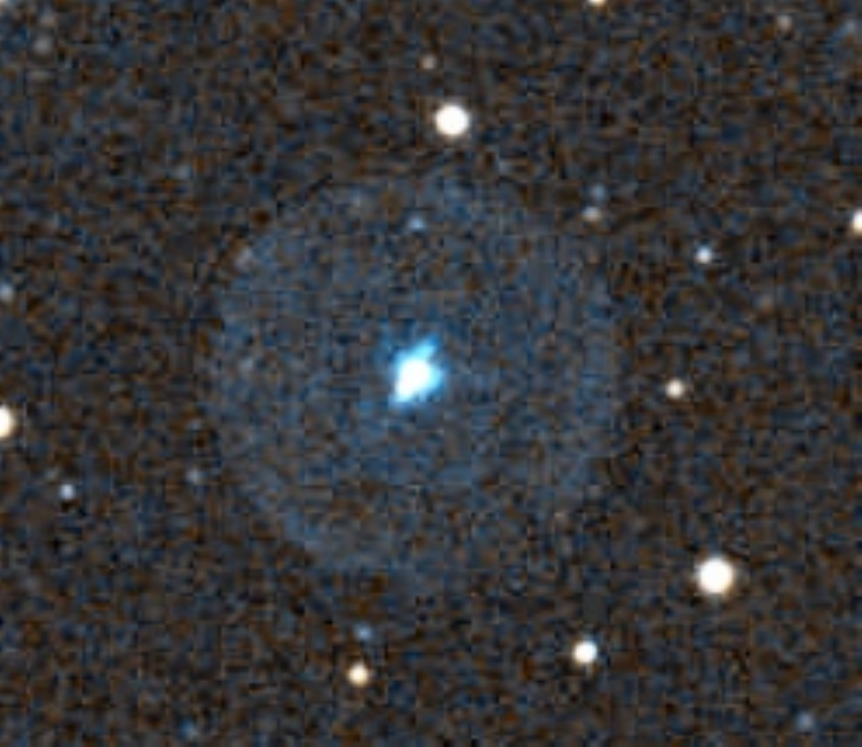}{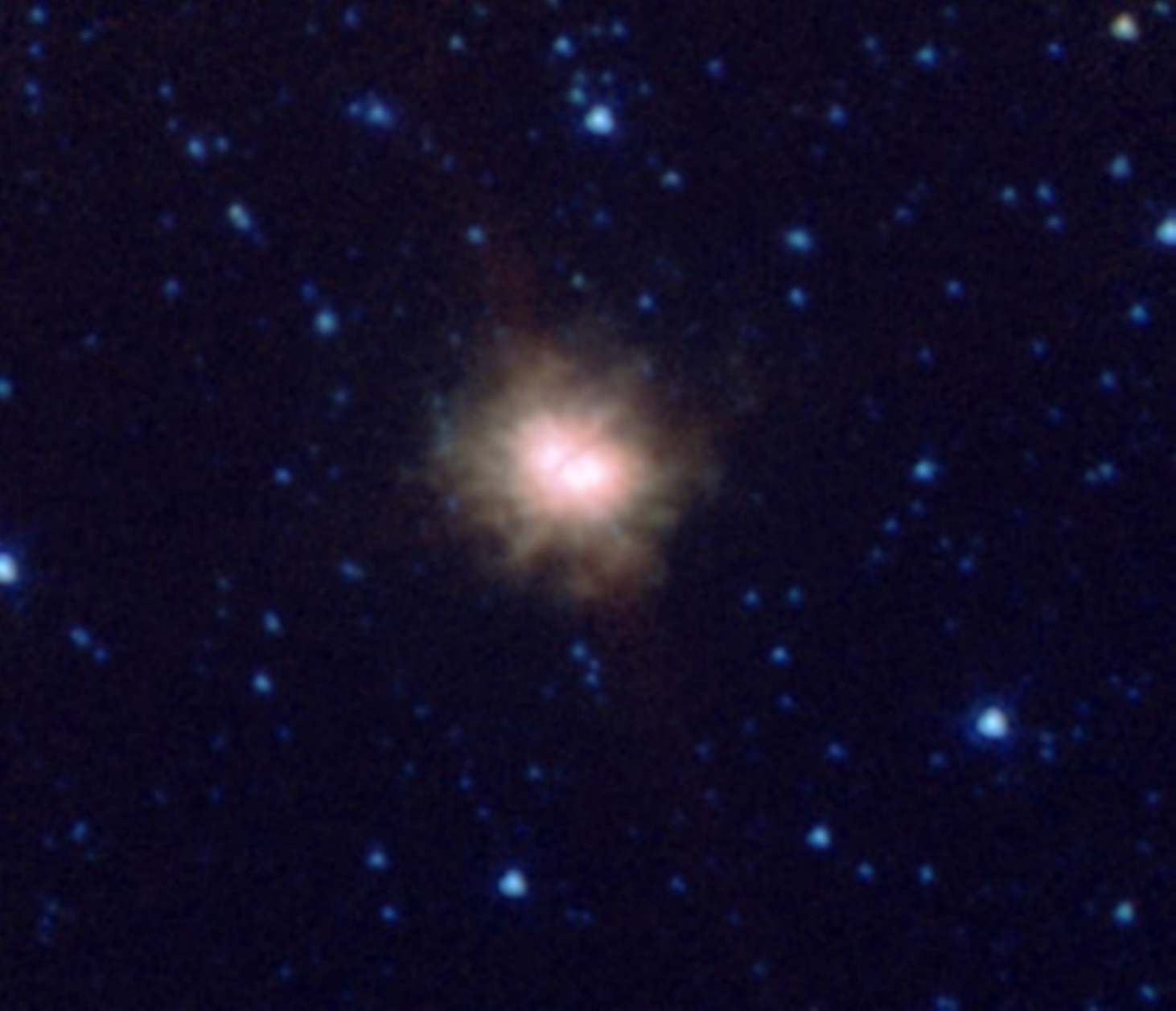}{}{Planetary Nebula Abell 30 in optical taken at Kitt Peak (left) and with IRAC (right). Optical image credit: Allan Cook/Adam Block/NOAO/AURA/NSF}

\section{Observations \& Results}
The sample of 80 targets were first observed with {\it Spitzer}; follow-up observations of 30 targets were performed using the Magellan I \& II Telescopes at Las Campanas Observatory in Chile. 

The three programs that were taken with {\it Spitzer} (Werner et al. 2004) that form the basis of this project were completed in 2009.  The sample comprises about 75 targets in total.  All have infrared spectra covering 5--40 $\mu$m$\,$ measured on the Infrared Spectrograph (IRS) (Houck et al. 2004) and were also observed with the Infrared Array Camera (IRAC) imager (Fazio et al. 2004), which images each object in four broadband channels centered at  3.6, 4.5, 5.8, and 8.0 $\mu$m$\,$ . 

The Magellan Clay telescope with the MMIRS (McLeod et al. 2004) was used to obtain near-infrared (H and K band) spectra of the observable subset of the {\it Spitzer} sample.   In addition, the Magellan Baade telescope was used to image the subset of sources in the {\it Spitzer} sample that have extended emission using near-Infrared narrowband filters that isolate $H_{2}$  and $Br \gamma$ emission with PANIC (Martini et al. 2004). 

Results of the Transition Object program have been published by Cerrigone et al. 2009. The imaging search (with IRAC and PANIC) for faint, extended emission around the post-AGB sample yielded a null result; SED and spectral analysis of the data is underway.  A sample IRAC image from the Planetary Nebula program is shown in Figure 1.  SED and spectral analysis is underway.

\bibliographystyle{asp2010}
\bibliography{apn5v1}

\end{document}